# Process Description, Behavior, and Control

Sabah Al-Fedaghi
Computer Engineering Department
Kuwait University
Kuwait
sabah.alfedaghi@ku.edu.kw

Haya Alahmad
Information Technology Department
Ministry of Public Works
Kuwait
haya.alahmad@ieee.org

*Abstract*—**Modeling processes** are the activities of *capturing* and *representing* processes and control of their dynamic behavior. Desired features of the model include capture of relevant aspects of a real phenomenon, understandability, and completeness of static and dynamic specifications. This paper proposes a diagrammatic language for engineering process modeling that provides an integration tool for capturing the static description of processes, framing their behaviors in terms of events, and utilizing the resultant model for controlling processes. Without loss of generality, the focus of the paper is on process modeling in the area of computer engineering, and specifically, on modeling of computer *services*. To demonstrate the viability of the method, the proposed model is applied to depicting flow of services in the Information Technology department of a government ministry.

## I. INTRODUCTION

A complex enterprise is established upon processes that include coordinated intermediate steps designed to *create*, *change*, *transfer*, and *receive* products and services. Complex systems rely on *modeling* (and simulation) methodologies to develop functional specifications, descriptions of flow of *things* (to be defined later), and system structure definitions [1]. This reliance on models may produce a multiplicity of *representations* that lead to confusion and difficulty in managing processes, as well as inconsistent usage [1].

*Process Modeling* (PM) involves [2]:
- Capturing and representing processes in the real world system
- Formalizing processes in preparation for such operations as automation.

Capturing and representing processes and control of their dynamic behavior produces abstract views at various levels of granularity of the system to be modeled. The model shows *how* a business case should be executed and managed. A conceptualization in this context refers to a depiction of the involved processes for use as a means of communication among stakeholders, like the blueprint of a house [3]. "It provides the basis of the model documentation; guides the development of a computer model; provides guidance for experiments; and is an aid for model verification and validation" [4]. Desired features of such a picture bring together relevant aspects of real phenomena (processes), understandability, and complete static dynamic specifications, and are independent of any implementation paradigm, e.g., software operations.

When we think of PM notations, we can identify a plethora of different approaches. This is due to the fact that during the historical development of process modeling notations, different communities have influenced the discipline of process modeling. [5]

In general, in business, engineering, and manufacturing processes, the abstract view is based on mathematical representation, or developed based on graphical languages [6]. The languages are tools for building conceptual models representing the static and dynamic aspects of a system that reflects a certain portion of reality. These languages include UML (and SysML) Activity Diagrams (ADs), Business Process Modeling Notation (BPMN), Event-driven Process Chains (EPC) [7], Specification and Description Language (SDL) [8], and Role-Activity Diagrams (RADs).

*A. Problem 1: contemporary modeling languages*

Nevertheless, current process models suffer from various weaknesses. "The first problem of contemporary modeling languages is their convertibility into machine language. This feature is not for free. In different words, they are derived from machine language and modeling a process in them means programming a process" [2]. This paper will not expand on such a broad topic; instead we offer a few examples in the context of BPMN and SysML.

According to Dijkman et al. [9],

> The mix of constructs found in BPMN makes it possible to obtain models with a range of semantic errors… [and] the static analysis of BPMN models is hindered by ambiguities in the standard specification and the complexity of the language.

This weakness has led to define BPMN in terms of Petri nets, for which efficient analysis techniques exist. "Thus, the proposed mapping not only serves the purpose of disambiguating the core constructs of BPMN, but it also provides a foundation to statically check the semantic correctness of BPMN models" [9].

SysML achieves only marginal success as a modeling tool in the development process because a multiplicity of fragmented representations in UML is exported to SysML, including narratives, diagrams, and notions [10]. Also, SysML lacks a nucleus from which different phases of the development process can evolve, analogous to lacking a blueprint in a complex building project such as a skyscraper, where it would be the core on which the building's framework, electrical system, water system, and interior walls are built by their various specialists [11].

*B. Problem 2: Conceptualizing basic notions*

In addition to difficulties raised by modeling languages, people have difficulty conceptualizing basic concepts in the field such as process, behavior, and events. To give an example, we focus on the current idea of an *event*. In BPMN, an event can do the following:

- appear at the beginning of a process, within a process (intermediate), or at the end of a process
- react to something, or "throw a result"
- be generic, time-based, message-based, rule-based, signal-based, exception-based, etc.
- be positioned within a sequence flow or attached to the boundary of an activity.
- interrupt current process execution or not.
- start a parallel, event-based sub-process. [12]

Conceptualizing a *start*, *intermediate,* and *end* of a process indicates that it has sub-processes, hence, the reason *only* three "pieces" of a complex process are declared. The ability to "react to something" seems to refer to a process that triggers another process. "Be positioned within sequence flow or attached at the boundary of an activity" gives the impression that an event comprises static descriptions of processes, while in actuality it is at the level of dynamic system behavior.

To illustrate a solution to this problem of the BPMN notion of event, we will demonstrate an alternative conceptualization of events.

*C. Contribution to the current problem*

This paper proposes another diagrammatic language, called Flowthing Machine (FM), for use in engineering process modeling. We claim that the resulting model provides a new integration tool (see Figure 1) for capturing a static description of processes (circle 1 in the figure), "eventizing" (identifying events in) their behaviors (2), and utilizing the resulting model to control the process.

Without loss of generality, this paper focuses on process modeling in the area of computer engineering; specifically, on modeling of computer *services*. Service-orientation is a paradigm that arose in the evolution of Information Technology that considers client needs and satisfaction the chief concern to be reflected in a quality *product*, including prompt response and sensitivity to client issues. As a demonstration of the viability of the method, the proposed model is applied to an actual flow of services in the Information Technology department of a government ministry.

FM has been utilized in several software engineering applications [13-21]. The next section briefly reviews some of its features. Section 3 discusses an example of the FM modeling approach, and Section 4 applies FM to our case study.

## II. FLOWTHING MODEL

The world is quite complex and includes objects, substances, actions, events, … as just a few of the *things* in this world. The FM Model views these things in term of flows that circulate through diverse fields, such as, for example, supply chain flows, money flows, and data flows in communication models. The basic construct in FM is a *thing*, represented in a diagrammatic (abstract) *machine* or *pattern* (see Figure 2). The machine represents a *process* (not to be confused with the process stage shown in Figure 2). Accordingly, the FM model represents a conceptual model of the real world that reflects all patterns of flow.

Typically (e.g., ISO 9000:2005) a process is defined as the transformation of inputs into outputs. Such a definition is incomplete [22] because "processes create results and not necessarily by transforming inputs" [22]. The claim in the Figure 2 representation is that *Transfer*, *Process*, *Release*, *Receive*, and *Create* are basic operations in any system, physical or otherwise.

*Things* in FM represent a range of physical and abstract items, including data, information, signals, objects, and events. The flow machine of Figure 2 is based on six stages (states), as follows:

**Arrive**: A thing reaches a new flow machine (e.g., data arrive at a buffer in a router machine).

**Accept**: A thing is permitted to enter a flow machine (e.g., it is addressed correctly for delivery); if arriving things are also always accepted, Arrive and Accept can be combined as a **Receive** stage.

**Release**: A thing is marked as ready to be transferred outside the flow machine (e.g., in an airport, airline passengers wait to board after passport clearance).

**Process** (change): A thing goes through some kind of transformation that changes its form but not its identity, e.g., a message can be translated into another language and a number can be changed from decimal to binary.

**Create**: A new thing is born (created) in a flow machine (e.g., a data mining program generates a conclusion).

**Transfer**: A thing is transported somewhere from/to outside the flow machine (e.g., packets reaching ports in a router, but still not in the arrival buffer).

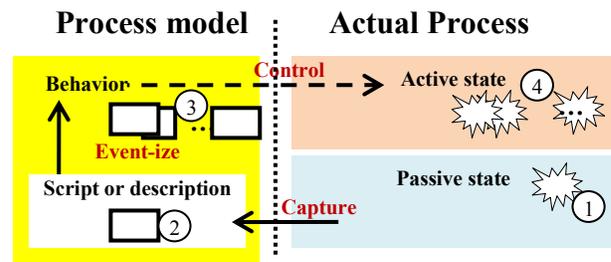

**Figure 1. Relationships between an actual process and its model in terms of description, behavior, and control**

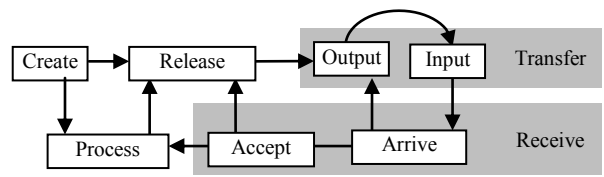

**Figure 2. Flow machine**

Note that typically a process is described with the essential components of input and output. In FM, input and output have a single conceptual function: *transfer*. It is possible that items that are output were not originally input, but rather were created internally. It is also possible that things that are input may not result in output.

In general, a flow machine is thought to be an abstract machine that receives, processes, creates, releases, and transfers things. The stages in this machine are mutually exclusive (i.e., a thing in the Process stage cannot be in the Create stage or the Release stage at the same time). An additional stage of *Storage* can also be added to any machine to represent the storage of flowthings; however, storage is not an exclusive stage because there can be stored processed flowthings, stored created flowthings, etc. Hereafter, a thing means a flowthing.

A flowthing is defined as a thing that can be created, released, transferred, arrived, accepted, or processed while flowing within and between machines. FM also uses the following notions:

*Spheres and subspheres*: These are the environments of the machine. Multiple machines can exist in a sphere if needed. A sphere can be an entity (e.g., a company, a customer), a location (a laboratory, a waiting room), a communication media (a channel, a wire). A flow machine is a subsphere that embodies the flow; it itself has no subspheres.

*Triggering*: Triggering is the creation or activation of a flow by a point or condition in another flow (denoted in FM diagrams by a **dashed arrow**); e.g., a flow of electricity triggers a flow of heat.

### III. APPLICATION TO PROCESSES

According to Vaillant [23], the following example was given as justification for the notion that "object oriented programming" is better than procedural programming.

*The school principal is standing in front of fresh students, who need to go to their respective classrooms. only the students don't know which classroom yet. how can we get each student to their assigned classroom? following object-oriented thinking, the principal solves the problem by telling each student where their assigned classroom is. "the focus is on the entities [objects]: the students and the principal. as objects, the students know how to handle themselves. they can speak their own name, and move by themselves.* [23]

This example will be used to illustrate the FM-based modeling and to develop a diagrammatic description, behavior, and control (management) of the scenario.

#### A. Functional description

For Umeda et al. [24], a *function* (represented by a schema) is "a description of behavior abstracted by human through recognition of the behavior in order to utilize it." FM separates development of a static description (structure) from modeling of behavior at a second level of specification using *events*. For now, we focus on a static description and will model behavior in the next subsection.

Figure 3 shows this static description using the FM representation. The *student sphere* includes:
- The flow (physical movement) of the student him/herself as a *thing* (circle 1)
- His/her classroom location data (2)

The *principal* (3) gives the student *data* (4) on the location of his/her class (5). The *Classroom* sphere (6) is the "system" that includes,
- Physical flow of the student (7).
- *Location data* (8) in possession of the student. Note that the movement of the student (red arrows) implies the flow of location data in his/her possession.
- The *actual location* (9) of the classroom is there as a "property", e.g., a sign hanging on the wall, a map indicating the location, someone informs about it, etc.
- A comparison happens (10) between the *location data* in the possession of a student with the *actual location* of the classroom, e.g., the student compares the class location data and physical location of the class, someone compares the data in the student hand with a sign on the wall, a computer reads the location data and announces that this place is the right/wrong class location, etc.

Accordingly, if the location data and the actual location are different (10), the student is triggered (11) to leave the classroom (12).

Figure 3 is a spatial blueprint of the region of *events* (to be discussed next) to take place as described in the given scenario. It is a *script* that has no dynamics of its own. Or it is the frame that constitutes the basin where events are set. It is "stillness" that is sublimated by the dynamism of events. "The world is certainly an ongoing process, but it can become an object of attention, learning, analysis, communication, and record only to the extent that such processes are apprehended and arrested in presumptively static forms" [25].

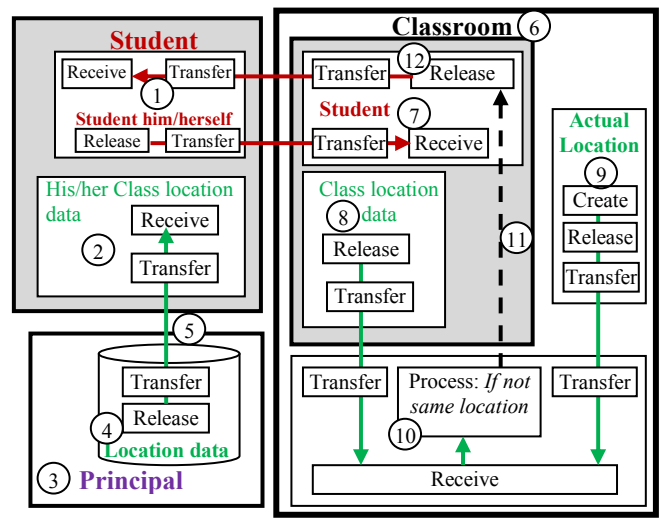

**Figure 3. FM representation of the example**

### B. Structure and behavior

The notions of *structure* and *behavior* have been explored in various engineering fields, where they play a central role:

> In recent years, … engineering design has placed a premium on being explicit and precise about many of the intuitive concepts related to design concepts such as function, *behavior*, *structure*, and causal relation. Today's systems and devices may have components from a number of domains. With the increase in such multidisciplinary design, … it is important to develop as general a framework as possible. [26] (Italic added)

Typically a system's *behavior* is associated with the notion of *state*, e.g., "a transition of states along time," where states consist of "entities, their attributes and their structure" [24]. It (behavior) "is represented by sequences of state transitions" [27]. According to Borgo et al. [28], the term *behavior* applied to "a technical artifact" refers to "the specific way in which the artifact occurs in an event."

In general, it is claimed that structure and behavior have many related meanings, and attempts to identify one true meaning are bound to fail [26].

Here *behavior* involves the behavior of things during *events* when the script (e.g., Figure 3) is acted upon. The chronology of activities can be identified by orchestrating the sequence of these events in their interacting processes. In FM, *an event is a thing that can be created, processed, released, transferred, and received*. A *thing* becomes active in events. Note that the process stage of an event means that an event *runs its course*. Accordingly, the choreography of the execution can emerge from the arrangement of events.

Modeling of behavior occurs in a phase that occurs after the structural description (e.g., Figure 3) and involves modeling the "events space" where an event is taking place or happening. The event is specified by its spatial area or subgraph (e.g., of Figure 3), its time, the event's own stages, and possibly by other things, e.g., such description of events as extent (strength). Note that a conceptual event refers to sets of (momentary or elementary) events extended in space and time that, in the context of the involved model, together form a meaningful event.

### C. Behavior description of the school principal

In Oxford Living Dictionaries, one definition of the verb *process* is to *perform a series of mechanical or chemical operations on (something) in order to change or preserve it*. A modified version of this definition suitable for our approach states that *to process is to perform a series of operations on (something) in order to change it*. The word process can also be stated as a noun meaning a *series of events that changes the states of a system*.

Combining this view with the discussion in the previous section, we can now identify the relationships between an actual process and its model as shown in Figure 1. In that figure, a *system* (e.g., student assignments to classrooms) can be viewed as a phenomenon in reality (1). It is described as a static script (2). The resulting description is "eventized": cut into pieces according to the natural joints of possible events. The resulting time-based schemata are used to *control and manage* execution of the system.

We can apply the FM concept of behavior to the school principal example. Figure 4 shows four selected "meaningful" events in the example:

**Event 1**: The principal gives a student his/her class location
**Event 2**: The student goes to the assigned location
**Event 3**: The student's assigned class location and the actual class location are compared.
**Event 4**: The student leaves the class if the two locations do not match

Accordingly, a procedure of execution for a student can be written as a sequence of events as follows:

(I) *Process: Event 1, Event 2, Event 3, if Event 4 then repeat Process*

Note that (I) represents *control* of the execution. *Control* here refers to directing the system's behavior or the course of *events*:

*Execute Event1, Execute Event 2, Execute Event 3, If Event 4 is executed then repeat the process.*
Generalizing the description,
*For all students*:
   {Event 1, Event 2, Event 3, if Event 4 then repeat}

In the example shown in Figure 5, a control level is applied to the school principal based on the relationships shown in Figure 1.

### D. Third level: Control

Now the total picture of the FM-based system appears. The functional and behavioral components have already been described. Typically, *control* is considered a mechanism to guide or regulate the behavior of a system so it functions as intended. "To control is to act, to put things in order to guarantee that the system behaves as desired" [29].

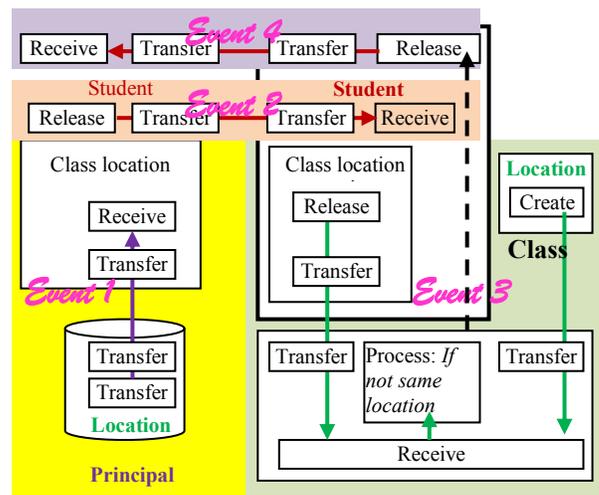

**Figure 4. Events**

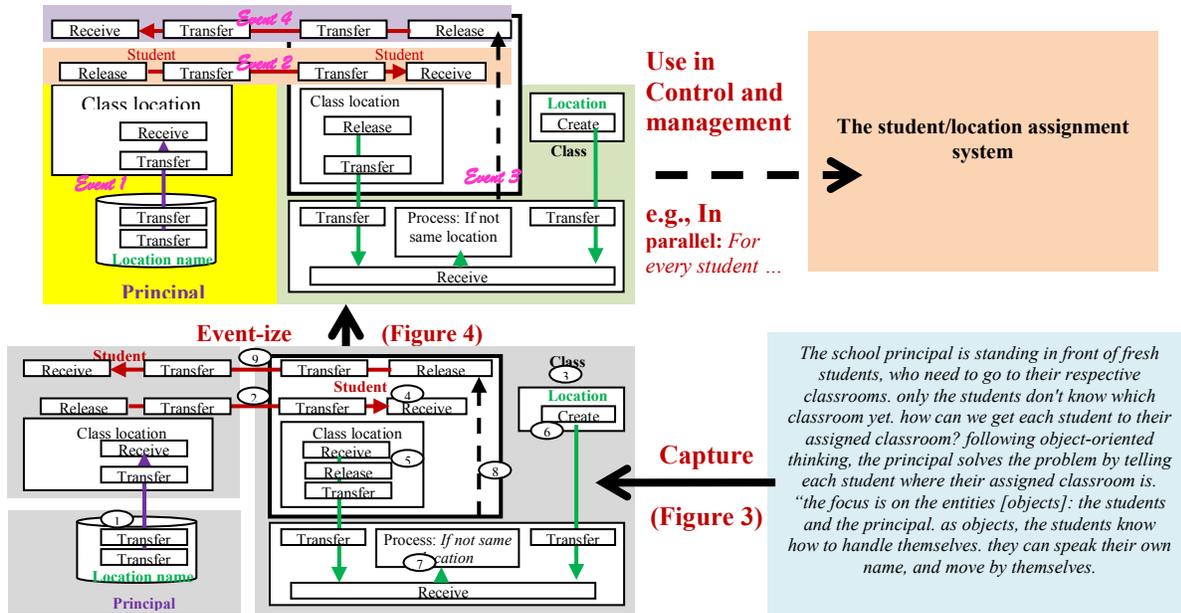

Figure 5. Illustration of the relationships between an actual process and its model

To limit the scope of the paper, especially since a new type of modeling is explored, we focus on the elementary "interior" control of an "error-free" system without monitoring and feedback. Hence, *control* here is limited to the configuration of facilitating system dynamics (events), e.g., sequence, parallelism, timing, etc. It is a simple control that oversees activities "according to specification." Control specification is usually viewed as the *behavior* of a system and includes a *state transition* diagram and *process activation* of behavior.

In FM, the control specification manages the activation of events, sequences, parallelism, etc. There are various types of such systems. For example, there is a mechanical system in which *control* involves merely turning the system on/off until its utility expires, e.g., a washing machine. There is a system controlled by another system without participating in the control decision, e.g., an automobile. Then there are all shades of systems that participate in the control function, e.g., feedback, self-managed systems, etc.

*E. Third level: Control*

The previous example utilizing FM methodology to integrate the description, behavior, and control of the process of assigning students to classes in one framework can be applied to all types of processes. As an example of FM used in the context of technical processes, consider the functional descriptions of an electrical buzzer given in [26].

*We ask the owner or the designer what the function of the device is. Consider the following answers:*
- *Buzzer-Function i. Its function is to make a sound come from box2, when the switch in box1 is closed.*
- *Buzzer-Function ii. Its function is to make box2 fill location2 with sound, when box1 is placed at location1 and box2 at location2, and when the switch in box1 is pressed. Etc. [26]*

A functional description of this electrical buzzer system is shown in Figure 6.

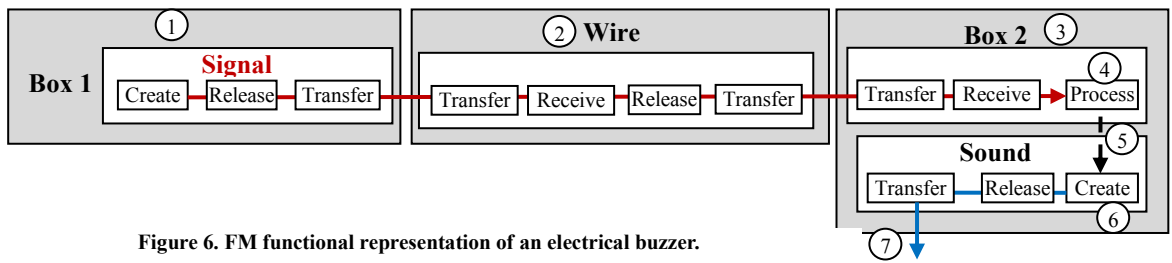

Figure 6. FM functional representation of an electrical buzzer.

Box 1 (circle 1) creates and sends a signal that flows to the wire (2) to reach Box 2 (3) where it is processed (4) to trigger (5) the generation (6) of sound to its surroundings (7). Figure 7 shows four "meaningful" events of an instance of the electrical buzzer.

According to Borgo et al. [28], "environment-centric meanings" can be extracted from meanings given by Chandrasekaran and Josephson [26], such as, for the electrical buzzer, "enabling a visitor to a house to inform the person inside the house that someone is at the door." In FM, this last meaning corresponds to the diagram shown in Figure 8. Here the visitor generates an action (1, pushing on Box 1) that triggers (2) creation of the signal. The signal causes Box 2 to generate sound that flows (3) to the person in the house (4) who processes it (5), triggering (6) generation of knowledge (7) that someone is at the door.

## IV. APPLICATION TO SERVICE PROCESSES IN AN IT DEPARTMENT

In the following section, we target process samples taken from the IT department of the Ministry of Public Works in Kuwait. Specifically we describe services provided by the network team such as Internet access and email. Because of space limitation, we describe here one sample service: facilitating access to the network based on privileges given by the administrators. In general, Figure 9 shows a snapshot of a basic network diagram in the ministry. It consists of multiple switches, routers, servers, and security elements.

Consider the description of the processes involved in providing network access to a new client, beginning with the two main service processes shown in Figure 10.

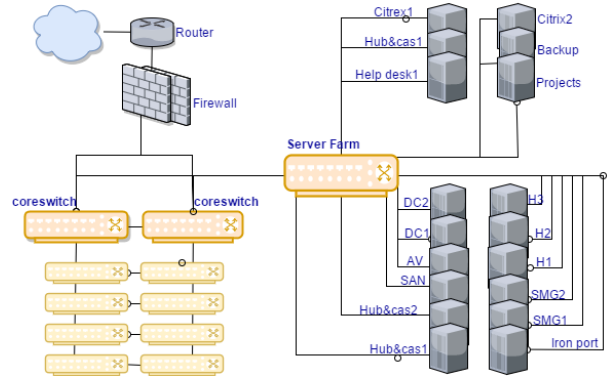

Figure 9. Snapshot of the network diagram of the IT department

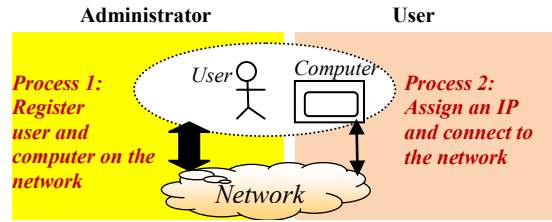

Figure 10. Assigning an IP to a new user on a specific computer

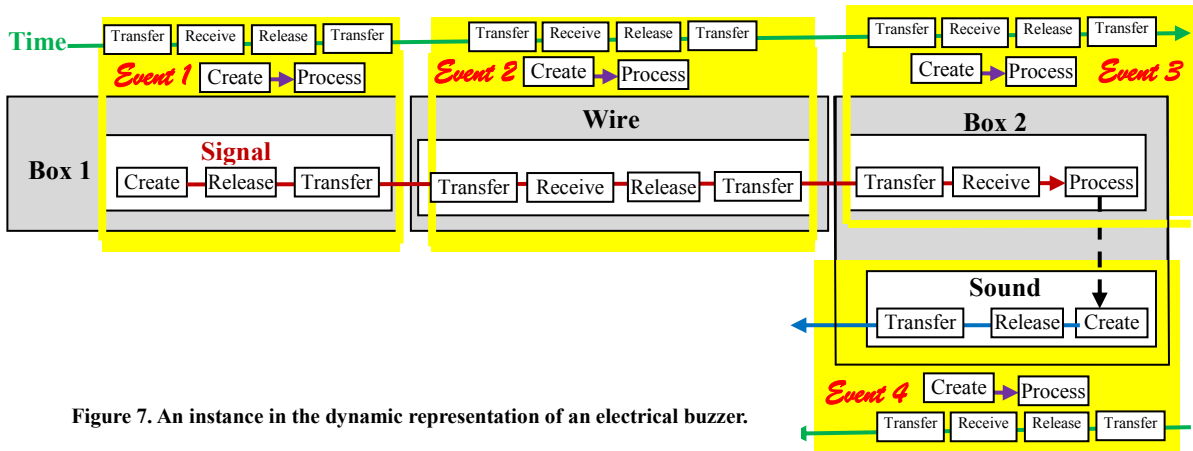

Figure 7. An instance in the dynamic representation of an electrical buzzer.

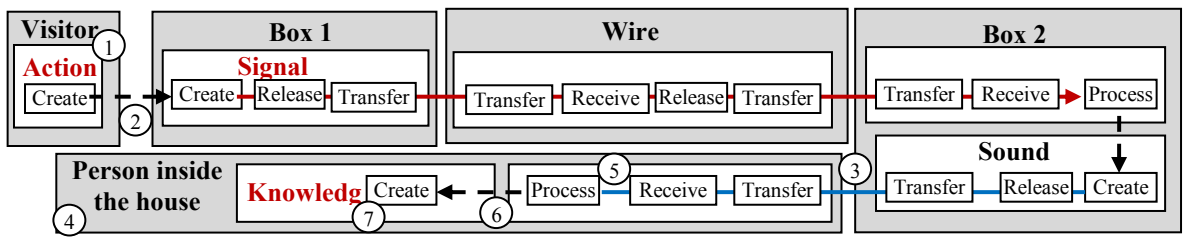

Figure 8. FM functional representation of the electrical buzzer

In current practice, these processes are described informally using text description along with some ad-hoc diagrams and general manuals not tailored to the local installation. This causes great difficulty in performing and maintaining these tasks and in documentation and training. The tasks can be described as follows.

*The system administrator performs the following tasks using his/her Active Directory:*
*Add New User Name*
*Assign Privileges to User*
*Delete User Name*
*Delete Computer Name*
*Then, the administrator performs the following task using the Client-computer:*
*Request to assign Computer-Name*
*At the end, the administrator gives the User his/her account and password*
*Accordingly, the User logs in for the first time, and this automatically creates a Request for Identity that is broadcast to all servers. The Dynamic Host Configuration Protocol (DHCP) responds by sending a TCP/IP package. The User then creates a Request to connect to Internet using this IP. Accordingly, the User is connected to the Internet.*

The conceptual description of these processes is necessary for the management and control of such tasks. The resulting high-level diagram plays the role of a network map for a network engineer. It can also be utilized by the help desk for control of its operations related to such tasks.

Figure 11 shows a static description of these two main processes in terms of the hardware and software architecture in use by the organization. The diagram can be understood as follows:

### A. The administrator registers a user on the network

Only the administrator has control over adding or removing a user from the network. Assume that the system administrator has received a request to assign an IP to a new user. The administrator (circle 1 in the figure) logs into his/her active directory and performs the following tasks:

**Adding New User Name**: A request to add a name is created (3) that flows (4) to Core Switch (5) and the network of switches to reach the server farm (6), then a domain controller server (7). There it is processed (8) and stored in the User-Name database.

**Assigning Privileges:** Similarly, the administrator assigns privileges to the user depending on the user's department and his/her job specification, following the flow path marked by circles 9, 10, 5, 6, 7, and 11.

**Assigning Computer-Name:** To designate the computer name to enable automatically linking the username with the computer name, the administrator logs into the user's computer (12) and creates a request (13). The request flows (14) through switches as before until it reaches (15) the Domain Controller (7), where the computer name is stored in a database of computer names (16).

Similar processes are followed for deleting user name (17) and computer name (18).

### B. Assigning an IP and connecting to the network

Accordingly, an account and its password are given to the user who logs in (19), automatically generating a request for Identity in the network IP (20). The request flows (21) through switches as before until it is broadcast to all servers accessible in the network (22), because the client machine still has no IP address to be identified with. The request is ignored by all servers except the DHCP (23) server, which processes the request (24) and responds by creating an IP (25) that flows back to the requester (26).

Accordingly, a request to connect to the web browser is generated (27) that includes an IP that flows through switches (28) until it reaches the IronPort (29 – a security server at the lower right corner of the diagram). Then it flows to one of the Domain Controllers (30) to be processed for authentication (31), flowing through a firewall (32) and followed by a router (33 - the gateway of the network) to reach the Internet (34). The router (33) also receives data from the network (34) that flow back to the user (35 – upper left corner of the diagram at the box labeled *request to connect to the web browser*) using the same path as the request to connect to the browser.

As described before, events can be identified and used in controlling processes including other specifications such as sequence, timing (*adding new users in patches on weekends*), constraints (e.g., *when security 2 alarm, Domain Controllers stop Transfer/receive from the Firewall*), policy, etc.

For example, consider two events:
- Event 1: The system administrator adds a new user and his/her privileges including the sub-event of assigning the computer name. This event is shown in Figure 12 (yellow in the online version of the paper).
- Event 2: The user logs in for the first time to receive the IP (orange in Figure 12).

The two events share the part of the system that communicates with related services (the domain controllers or the Dynamic Host Configuration Protocol). This part of the system appears in purple.

Note that these two events include several levels of sub-events. Such a diagrammatic identification of regions of events can be used to specify different types of constraints at the control level. For example, it is possible to specify that *the user should log in to receive IP within t (e.g., a week) from registering in the system, otherwise the account is deleted*. That is, *{Event 1, Event 2} < t*.

## V. CONCLUSION

This paper has proposed a diagrammatic language for engineering process modeling that provides an integration tool for capturing the static description of processes, identifying the events in their behaviors, and utilizing the consequent model in controlling processes.

**Figure 11. Description of the processes of registering a user and his/her computer on the network, assigning an IP, and connecting to the network**

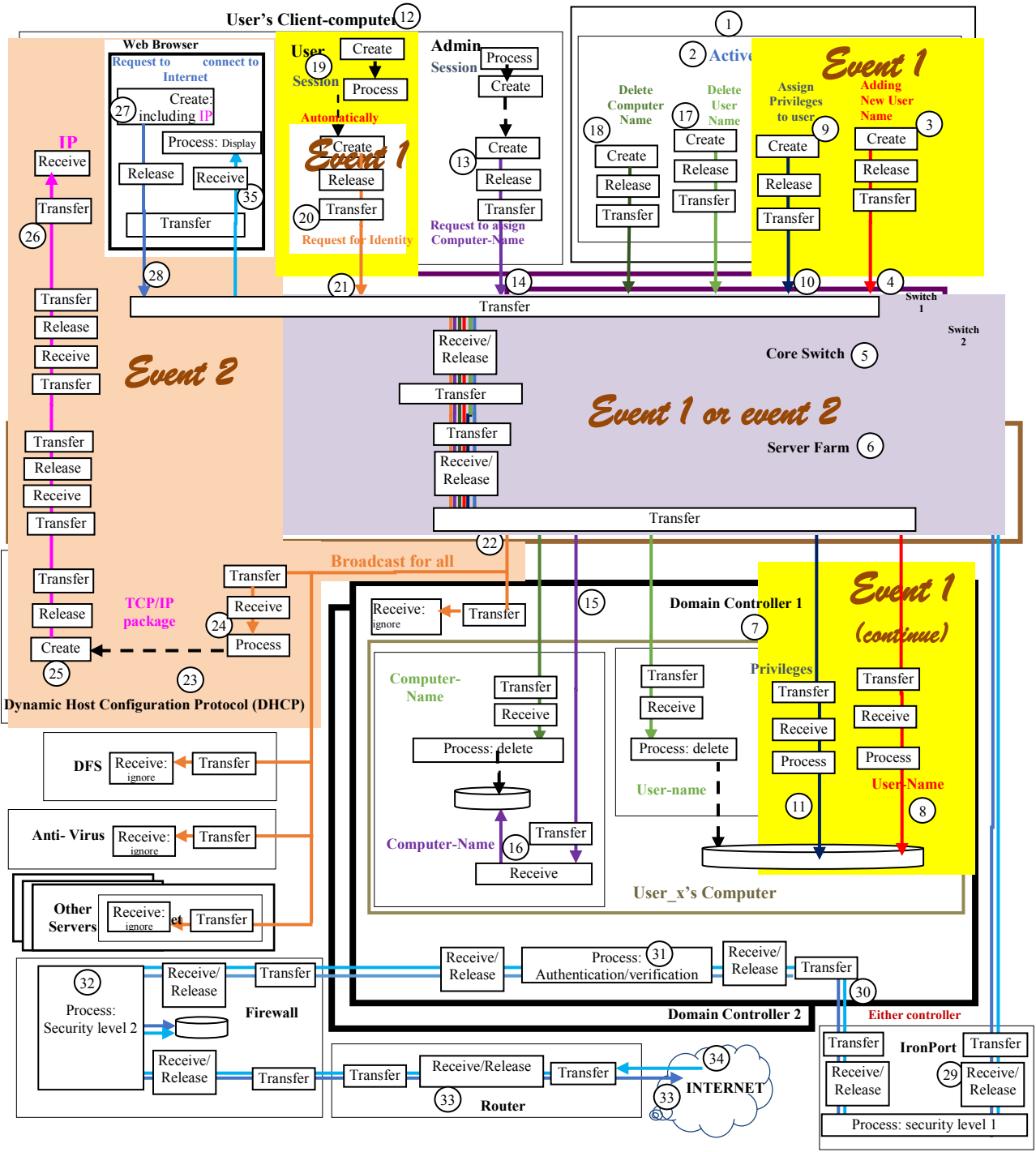

**Figure 12. The event of registering a new user and his/her computer in the network and the event of assigning an IP and connecting to the network**

The study has applied the proposed model to an actual process sample taken from the IT department of a government ministry. The resulting description demonstrates the viability of the methodology that can be adapted to different types of processes. Further research will reveal the applicability of the modeling technique to real systems.

AUTHORS PROFILE

Sabah Al-Fedaghi holds an MS and a PhD in computer science from the Department of Electrical Engineering and Computer Science, Northwestern University, Evanston, Illinois, and a BS in Engineering Science from Arizona State University, Tempe. He has published two books and more than 270 papers in journals and conferences on software engineering, database systems, information systems, computer/ information privacy, security and assurance, information warfare, and conceptual modeling. He is an associate professor in the Computer Engineering Department, Kuwait University. He previously worked as a programmer at the Kuwait Oil Company and headed the Electrical and Computer Engineering Department (1991–1994) and the Computer Engineering Department (2000–2007).

Haya Alahmad holds Bachelor's and Master's degrees in computer engineering from the Department of Computer Engineering, Kuwait University. She worked in Zain Telecom company from 2013 to 2014. She has been working since 2015 as a computer engineer in the Information Technology Department, Ministry of Public Works, Kuwait. Her interests include computer networks and software engineering.